# Nonsymmorphic nodal-line metals in the two-dimensional rare earth monochalcogenides MX (M = Sc, Y; X = S, Se, Te)

Hao Guo[1,2], Jianzhou Zhao[2,3*], Cong Chen[2,4], Si Li[2,5], Wentao Jiang[1], Xiaobao Tian[1,†], Shengyuan A. Yang[2]

[1] *Department of Mechanics and Engineering, Sichuan University, Chengdu, Sichuan 610065, China*

[2] *Research Laboratory for Quantum Materials, Singapore University of Technology and Design, Singapore 487372, Singapore*

[3] *Co-Innovation Center for New Energetic Materials, Southwest University of Science and Technology, Mianyang, Sichuan 621010, China*

[4] *Key Laboratory of Micro-nano Measurement-Manipulation and Physics (Ministry of Education), School of Physics, Beihang University, Beijing 100191, China*

[5] *Key Laboratory of Low-Dimensional Quantum Structures and Quantum Control of Ministry of Education, Department of Physics and Synergetic Innovation Center for Quantum Effects and Applications, Hunan Normal University, Changsha 410081, China*

**Abstract**

We predict a new family of two-dimensional (2D) rare earth monochalcogenide materials $MX$ ($M$ = Sc, Y; $X$ = S, Se, Te). Based on first-principles calculations, we confirm their stability and systematically investigate their mechanical properties. We find that these materials are metallic and interestingly, they possess nodal lines in the low-energy band structure surrounding the whole Brillouin zone, protected by nonsymmorphic crystal symmetries in the absence of spin-orbit coupling (SOC). SOC opens small energy gaps at the nodal line, except for two high-symmetry points, at which fourfold degenerate 2D spin-orbit Dirac points are obtained. We show that these topological band features are robust under uniaxial and biaxial strains, but can be lifted by the shear strain. We also investigate the optical conductivities of these materials, and show that the transformation of the band structure under strain can be inferred from the optical absorption spectrum. Our work reveals a new family of 2D topological metal materials with interesting mechanical and electronic properties, which will facilitate the study of nonsymmorphic symmetry enabled nodal features in 2D.

---

* jianzhou_zhao@sutd.edu.sg
† xbtian@scu.edu.cn

# 1. Introduction

Topological metals/semimetals have attracted broad interest in multiple fields of physics, chemistry, and materials science, owing to their fascinating properties and potential applications in many frontiers of research [1-7]. In such materials, the electronic bands form robust degeneracies around the Fermi level, such that the low-energy electrons could have unusual dispersion as well as emergent pseudospin degree of freedom, leading to their exotic behavior. Such degeneracies may take the form of nodal points [8-12], nodal lines [13-18], or even nodal surfaces [19-21]. And their robustness derives from the topology/symmetry protections. Based on the type of protection, the band nodal features can be distinguished into two classes. The first class is called accidental, because these nodal features can be removed by a continuous band deformation without changing the symmetry of the system. The other class is known as the essential degeneracies, which cannot be removed by any symmetry-preserving deformation. The accidental degeneracies are typically protected by symmorphic symmetries, such as rotation or reflection, whereas the essential degeneracies require nonsymmorphic crystal symmetries, i.e., screw rotation or glide reflection, which involve fractional lattice translations [9,22,23]. In three dimensions (3D), a range of candidate materials have been identified for each class.

When moving to the two dimensional (2D) world, the number of symmetry operations are greatly reduced. As a result, the discovered 2D topological metals are also much less. This is especially the case for the class with essential degeneracies [24], by noting that in 2D, we have only one glide mirror and two screw axes, all lying within the 2D plane. So far, the reported nonsymmorphic 2D topological materials are quite limited. Examples include the spin-orbit Dirac points in HfGeTe-family monolayers [25], 2D $X_3SiTe_6$ ($X$ = Ta, Nb) [26], and α-bismuthene [27]; hourglass nodal loops in the GaTeI family monolayers [28]; and magnetic nodal lines in monolayer CoSe [29]. And the only experimental verification was on the 2D spin-orbit Dirac point in α-bismuthene [27]. Thus, to facilitate the theoretical and experimental studies of these novel topological states, there is an urgent need to identify new 2D materials with nonsymmrophic band degeneracies.

In this work, based on the first-principles calculations, we predict a new family of 2D

materials, the 2D rare earth monochalcogenide *MX* (*M* = Sc, Y; *X* = S, Se, Te), and we show that they possess nonsymmorphic nodal line and nodal points nears the Fermi level. Our study is motivated by noting that in the 3D bulk form, these materials take a cubic NaCl-type structure. However, when reduced to a 2D single layer (corresponding to a bilayer in the bulk), the structure will have a spontaneous distortion and adopt a wrinkled structure. Via first-principles calculations, we confirm that these 2D structures are stable and we systematically investigate their mechanical properties. Importantly, such a structural change also modifies the symmetry: there emerge two in-plane screw axes, making the space group nonsymmorphic. We show that these nonsymmorphic symmetries give rise to a nodal line near the Fermi level in the absence of spin-orbit coupling (SOC), surrounding the whole Brillouin zone (BZ). SOC opens a small gap on the nodal line, except for two high-symmetry points, where fourfold degenerate spin-orbit Dirac points are realized. We investigate the change of the band structure under various lattice strains, and demonstrate that the nodal line is robust against the biaxial and uniaxial strains, but is lifted by the shear strain. Furthermore, we study the optical conductivity of these materials, and show that the change in the nodal line may be traced by measuring the optical absorption spectrum. Finally, we screen the commonly used substrate materials, and suggest $SrTiO_3$ and $TiO_2$(001) as the suitable substrates for the growth of these materials.

## 2. Computational methods

Our first-principles calculations were performed by using the Vienna ab initio Simulation Package (VASP) [30, 31] based on the density functional theory (DFT). The projector augmented wave (PAW) pseudopotentials [32] were adopted. The exchange-correlation functional was described within the general gradient approximation (GGA) with the Perdew, Burke and Ernzerhof (PBE) realization [33]. A vacuum of 30 Å was added to eliminate the artificial interaction between periodic images. The cutoff energy was set as 500 eV, and the first BZ was sampled using a Γ-centered *k*-point mesh of size 13×13×1 [34]. The geometric structures were fully relaxed with the force and energy convergence criteria set as 0.01 eV/Å and $1.0×10^{-6}$ eV, respectively. Phonon spectra were calculated using the frozen phonon method and generated with the PHONOPY package [35]. The optical conductivities were calculated using the WIEN2K package [36]. The plane wave cut-off parameter $K_{max}$ was given by $R_{mt}*K_{max}$ = 8.0. The $k$-point mesh for the self-consistent loop was the same as the VASP input, and it was increased to 100×100×1 for optic conductivity calculation.

## 3. Results

### 3.1. Lattice structure

The bulk rare earth monochalcogenides $MX$ ($M$ = Sc, Y; $X$ = S, Se, Te) share the NaCl-type cubic crystal structure (*Fm3-m*, No. 225) (which can be transformed to the CsCl-type structure (*Pm3m*, No. 221) under pressure) [37-40]. These materials have excellent thermal stability and they have been enjoying many practical applications in the fields of nonlinear optics, electro-optic devices, glass-making, grinding alloys, composites lasers, phosphors, and electronics [41, 42]. In this work, we explore the 2D form of these materials. We take a unit of two atomic layers from the bulk NaCl-type cubic structure. After relaxation, we find that the optimized 2D $MX$ (referred to as the monolayer $MX$ in the following), although maintaining a square lattice in the 2D plane, becomes corrugated in the $z$-direction, as shown in Fig. 1(a) and 1(b). The space group symmetry is transformed to *P4/nmm* (No. 129) by the corrugation, and is the same as the 2D Pb$X$ ($X$ = S, Se, and Te) monolayers [43, 44]. We note that a previous study on the superconductivity in 2D YS had also predicted this structure [45].

In their 2D structure, there are two $M$ atoms and two $X$ atoms per unit cell. The optimized lattice constants and the bond lengths of these $MX$ monolayers are listed in Table 1. The lattice has inversion symmetry $\mathcal{P}$. Importantly, the space group is nonsymmorphic: there exist two screw axes in the 2D plane, $\tilde{S}_x: (x, y, z) \rightarrow (-x + 1/2, y + 1/2, -z)$ and $\tilde{S}_y: (x, y, z) \rightarrow (x + 1/2 - y + 1/2, -z)$. Besides, there is no magnetic ordering found in these materials. Hence, the time reversal symmetry $\mathcal{T}$ is also preserved. These symmetries will be essential for our discussion below.

To confirm the dynamical stability of the obtained structures, we calculate their phonon spectra (plotted in the Supporting Information Fig. S1). As a representative, Fig. 1(d) shows the result of monolayer ScTe. Obviously, there is no imaginary frequency (soft mode) in the whole BZ, indicating that the structure is dynamically stable. There are 12 vibrational modes appearing in the spectrum. The acoustic ZA branch with a quadratic dispersion, which is a characteristic for 2D materials [46, 47], can be clearly observed here.

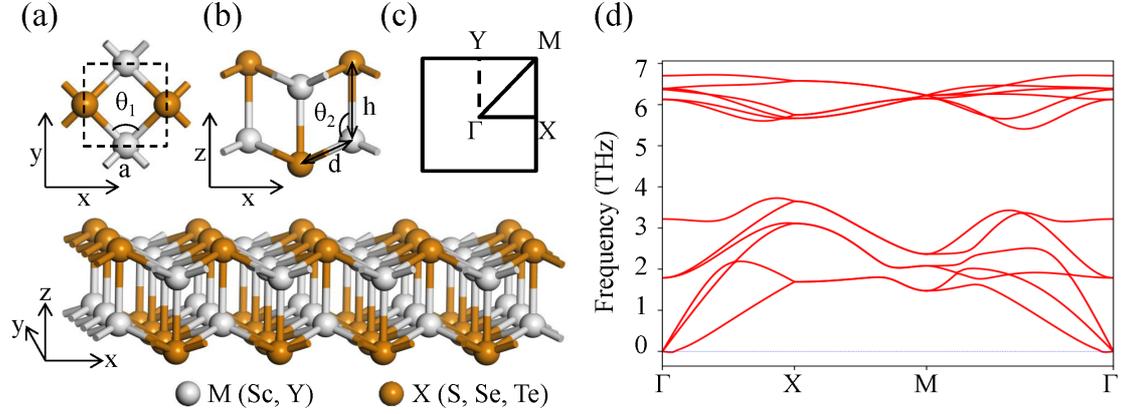

Fig. 1 (a) Top view and (b) side view of the MX monolayers. The grey and orange atoms represent X (S, Se, Te) and M (Sc, Y) atoms, respectively. The lattice constant a, the bond lengths d and h, and the bond angles $\theta_1$ and $\theta_2$ are labeled in the figure. (c) The first BZ with the high symmetry points. (d) Phonon spectrum of the monolayer ScTe.

Table 1 Calculated lattice parameters of MX monolayers, including the lattice constant (a), bond lengths (d and h), and bond angles ($\theta_1$ and $\theta_2$).

| MX | Lattice constant (Å) | Bond length (Å) | | Bond angle (deg.) | |
|---|---|---|---|---|---|
| | a | d | h | $\theta_1$ | $\theta_2$ |
| ScS | 3.571 | 2.581 | 2.616 | 87.558 | 101.922 |
| ScSe | 3.709 | 2.713 | 2.753 | 86.247 | 104.867 |
| ScTe | 3.937 | 2.929 | 2.949 | 84.447 | 108.153 |
| YS | 3.816 | 2.744 | 2.790 | 88.093 | 100.556 |
| YSe | 3.951 | 2.870 | 2.926 | 86.991 | 103.266 |
| YTe | 4.170 | 3.076 | 3.126 | 85.342 | 106.560 |

## 3.2. Mechanical properties

The mechanical properties provide information on the stability and stiffness of a material. Here, we have calculated the elastic constants ($C_{ij}$), Young's moduli (Y), and Poisson ratios (ν) of the MX monolayers. The results are listed in Table 2. The standard Voigt notations are used, with 1-xx, 2-yy, and 6-xy. From the results, we confirm that the elastic constants fulfill the mechanical stability criterion of 2D materials with square lattices: $C_{11}C_{22} - C_{12}^2 > 0$ and $C_{66} > 0$ [48] ($C_{22} = C_{11}$ for the square lattice), indicating that the *MX* monolayers are mechanically stable. The obtained Young's moduli range from 61.491 N/m ~ 104.075 N/m, which are smaller than their bulk materials (e.g., in the bulk, ScS: 141.43 N/m; ScSe: 114.04 N/m; ScTe: 82.76 N/m) [42] and 2D materials such as graphene (340 ± 50 N/m) [49], monolayer $MoS_2$ (120 N/m) [50] and monolayer BN (267 N/m) [51]. Meanwhile, their Poisson ratios range from 0.356 ~ 0.456, which are larger than graphene (0.178) [52], monolayer $MoS_2$ (0.254) [50], monolayer BN (0.21), and SiC (0.29) [51]. These results imply that the *MX* monolayers are less stiff and have better flexibility. Therefore, they should be more susceptible to strain engineering, as we will discuss in a while.

Table 2 Calculated elastic constants ($C_{ij}$), Young's moduli (Y) and Poisson ratios (ν) of *MX* monolayers. The standard Voigt notations are used: 1-xx, 2-yy and 6-xy.

| MX | Elastic constants (N/m) | | | Young's moduli (N/m) | Poisson's ratio |
| --- | --- | --- | --- | --- | --- |
| | $C_{11}$ | $C_{12}$ | $C_{66}$ | Y | ν |
| ScS | 121.648 | 46.247 | 65.120 | 104.075 | 0.380 |
| ScSe | 102.676 | 36.832 | 53.583 | 89.467 | 0.359 |
| ScTe | 79.587 | 28.301 | 40.244 | 69.522 | 0.356 |
| YS | 114.136 | 52.025 | 66.544 | 90.429 | 0.456 |
| YSe | 97.600 | 44.552 | 57.851 | 77.265 | 0.456 |
| YTe | 76.489 | 33.884 | 46.544 | 61.491 | 0.443 |

**3.3. Electronic structures and nonsymmorphic nodal lines**

Now, we turn to the electronic properties of the *MX* (*M* = Sc, Y; *X* = S, Se, Te) family materials. As they share very similar features, in the following, we take ScTe as a representative for presentation. The results for other members are relegated to the Supporting Information. The band structure and the partial density of states (PDOS) of monolayer ScTe without SOC are shown in Fig. 2. The result shows that the material is metallic. From PDOS, one observes that the low-energy states are mainly contributed by the *d* orbitals of Sc and *p* orbitals of Te. To analyze the bonding character, we have also calculated the electron localization function (ELF) (presented in Supporting Information Fig. S3), which indicates that the material is dominated by ionic bonding.

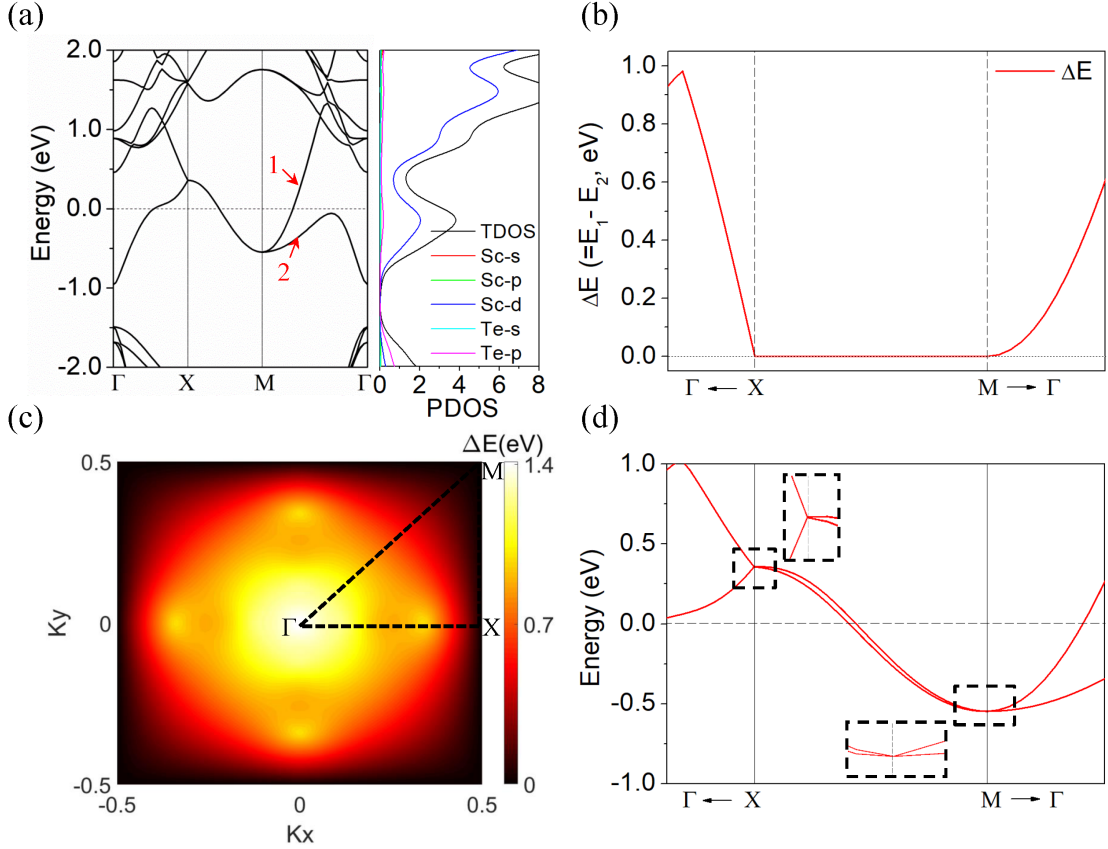

Fig. 2 (a) Band structure and PDOS of monolayer ScTe in the absence of SOC. The Fermi level is set at zero. (b) The energy difference ($\Delta E = E_1 - E_2$) curve for the two bands labeled as 1 and 2 in (a). (c) The 2D image of the energy difference $\Delta E$ in the whole BZ. (d) Band structure near the Fermi level for monolayer ScTe when SOC is included. The inserts show the enlarged views around the X and M points.

Interestingly, from the band structures in Fig. 2(a), one notes that there are two energy bands (labeled as 1 and 2) which linearly cross along the high-symmetry path X-M around the Fermi level. To confirm this band degeneracy, we plot the energy difference $\Delta E$ ($\Delta E = E_1 - E_2$) along the X-M path [Fig. 2(b)] and in the whole BZ [Fig. 2(c)]. The results show that these two energy bands are indeed degenerate along the boundary of the whole BZ, indicating that monolayer ScTe is a nodal-line metal in the absence of SOC.

Next, we investigate the mechanism for the protection of the nodal line. Focusing on the X-M path with $k_x = \pi$ in Fig. 2(c), we note that any $k$ point on this path is invariant under a combined symmetry operation $\mathcal{T}\tilde{S}_x$. This operation satisfies

$$(\mathcal{T}\tilde{S}_x)^2 = T_{100} = e^{-ik_x}$$

where $T_{100}$ represents a translation along the [100] direction by one unit cell. Here, we have $\mathcal{T}^2 = 1$ for a system without SOC (as for such case, the electrons can be regarded as spinless). Along the X-M path, $k_x = \pi$, hence $T_{100} = -1$. Therefore, we have $(\mathcal{T}\tilde{S}_x)^2 = -1$ on this path. This anti-unitary operation thus generates a Kramers-like double degeneracy for every point on X-M. Deviating from the path, the protection is lost, hence the two bands will generally split. This gives rise to the nodal line along X-M. Furthermore, because the boundaries of the BZ are connected by the fourfold rotation along $z$, the nodal line occurs for the whole BZ boundary.

SOC generally lifts the degeneracy at the nodal line. From Fig. 2(d), one can observe that SOC opens a small gap for the nodal line on the X-M path. Nevertheless, the degeneracy at the two points X and M is maintained. Here, after including SOC, each band has a twofold spin degeneracy due to the $\mathcal{PT}$ symmetry. Hence, the crossing points at X and M are fourfold degenerate Dirac points. Moreover, such Dirac points are robust under SOC (in comparison, the nodal point in graphene is removed by SOC), so they belong to the 2D spin-orbit Dirac points discussed in Ref. [25, 26].

Below, we clarify the symmetry protection of these spin-orbit Dirac points. In the analysis, we will utilize the glide mirror $\tilde{M}_z : (x, y, z) \rightarrow \left(x + \frac{1}{2}, y + \frac{1}{2}, -z\right)$ and the mirror $M_y$ instead of the screw rotation. These mirror symmetries can be obtained from the twofold (screw) rotations by combining with the inversion $\mathcal{P}$.

Consider the Dirac point at X $(\pi, 0, 0)$. This point is invariant under the symmetries $M_y$, $\tilde{M}_z$, and $\mathcal{T}$. Each energy eigenstate at X can also be chosen as an eigenstate of $M_y$. Since

$$(M_y)^2 = \bar{E} = -1,$$

the $M_y$ eigenvalues must be $m_y = \pm i$. For a Bloch state $|u\rangle$ at X with an $M_y$ eigenvalue $m_y$, its Kramers partner $\mathcal{T}|u\rangle$ must have the eigenvalue $-m_y$. Meanwhile, the commutation relation between $M_y$ and $\tilde{M}_z$ is given by

$$M_y \tilde{M}_z = -T_{0\bar{1}0} \tilde{M}_z M_y,$$

where the minus sign comes from the anticommutativity between the mirror operations on spin. At X, we have

$$M_y(\tilde{M}_z \mathcal{T}|u\rangle) = m_y(\tilde{M}_z \mathcal{T}|u\rangle),$$

which shows that the two orthogonal states $|u\rangle$ and $\widetilde{M}_z\mathcal{T}|u\rangle$ have the same $m_y$. Therefore, the following four states $\{|u\rangle, \mathcal{T}|u\rangle, \widetilde{M}_z|u\rangle, \widetilde{M}_z\mathcal{T}|u\rangle\}$ are linearly independent and degenerate with the same energy, forming the spin-orbit Dirac point at X. The Dirac point at M can be argued in a similar way.

We have a few remarks before proceeding. First, we have demonstrated that the nonsymmorphic symmetries play a crucial role in stabilizing the nodal line and the spin-orbit Dirac point here. These nodal features are guaranteed to exist by the symmetry. Hence, they belong to the essential band degeneracies.

Second, the protection of the nodal line by the $\mathcal{T}\widetilde{S}_x$ symmetry is analogous to the protection of so-called Class II nodal surfaces in 3D systems [21]. For both cases, the nodal feature must appear at the boundary of the BZ. Particularly, the previous study had shown that the protection works in the absence of SOC. In the presence of SOC, the nodal feature can still remain if the $\mathcal{PT}$ symmetry is broken [21]. For the monolayer *MX* considered here, the $\mathcal{PT}$ symmetry is preserved, hence the nodal line will be lifted and reduced to the Dirac points.

Third, it should be pointed out that the splitting of the nodal line by SOC is small. The result shown in Fig. 2(d) is for ScTe. For other members with lighter elements, the SOC strength (and hence the splitting) is even smaller, on the order of few meV. Consequently, regarding most measurable properties, the SOC effect can be neglected for these materials, and one can consider these materials as nodal-line metals (we have explicitly verified this point in the optical conductivity calculation, see Supporting Information). In the following, we will neglect SOC in the analysis.

### 3.4. Strain effects and optical conductivity

Lattice strain has been proved to be an effective method to tune the physical properties of 2D materials [53-56]. Below, we explore the effects of different types of strains on the properties of *MX* monolayers.

We first consider the uniaxial and biaxial strains, as schematically illustrated in Fig. 3(a). Figure 3(b) shows the calculated stress-stain curves for monolayer ScTe. The results show that the material is quite flexible. The critical uniaxial and biaxial tensile strains can reach ~ 30% and 17%, respectively. The linear elastic regime can be up to about 6% strain.

Figure 3(c) and 3(d) respectively show the change of the band structure under uniaxial and biaxial strains of 10%. One can observe that the change is not much. Importantly, the nodal line is preserved under both strains. The strains mainly make the nodal line band flatter and closer to the Femi level. This robustness of the nodal line can be understood by noting that the two screw axes $\tilde{S}_x$, $\tilde{S}_y$ and the time-reversal symmetry are preserved by the strain. Thus, the nodal line is still protected and remains an essential band degeneracy.

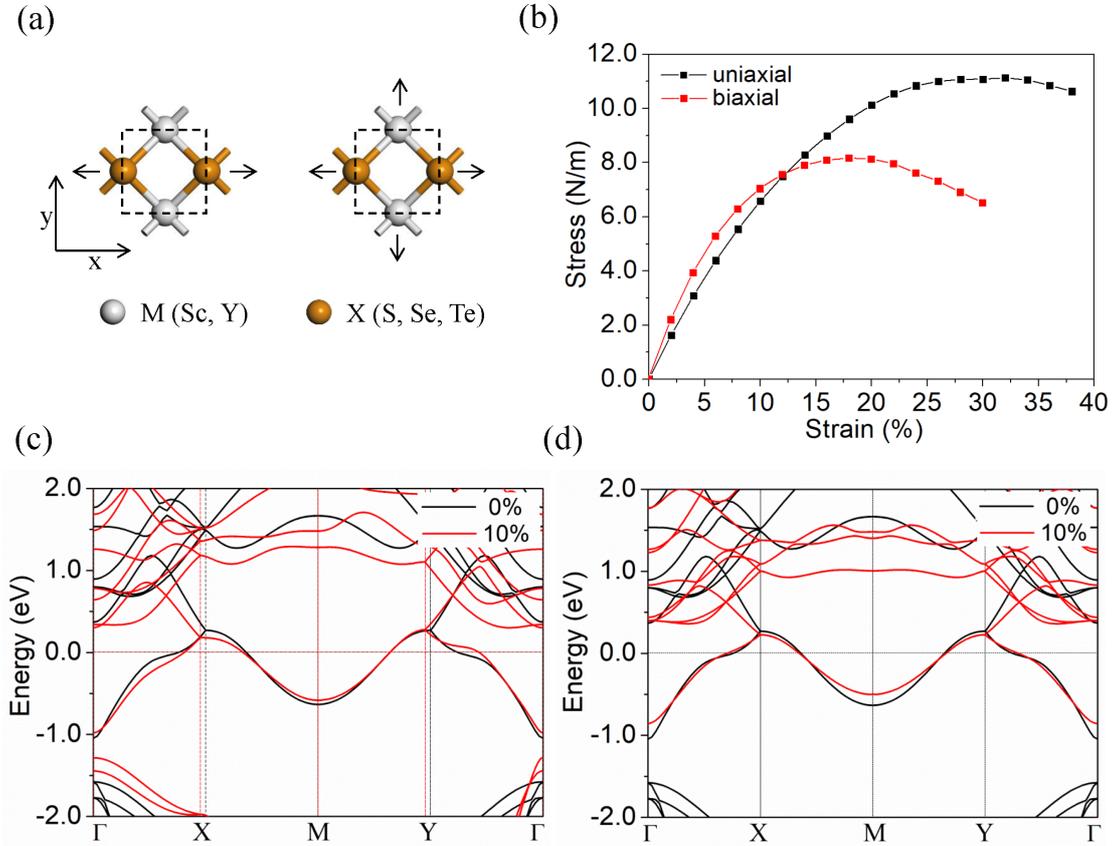

Fig. 3 (a) Schematic diagrams showing the uniaxial and biaxial strain acting on the 2D structure. (b) Calculated stress-strain curves for monolayer ScTe. (c, d) The change in the band structure for monolayer ScTe under (c) 10% uniaxial strain and (d) 10% biaxial strain.

However, the screw axes may be broken by the shear strain. Here, we consider the type of shear strain by varying the angle γ as shown in Fig. 4(a) from its equilibrium value of 90 degree. Under this strain, the lattice symmetry is changed to the space group *Cmma*. From the calculated band structure in Fig. 4(b), one confirms that the nodal line no longer exists, because the original screw rotational symmetry is broken. This result again illustrates the important role played by the nonsymmorphic symmetry in stabilizing the nodal line.

To probe the band structure and the effect of strain, we consider the optical conductivity of monolayer *MX*. Optical conductivity determines the linear response of a material to oscillating electromagnetic fields. In experiment, it can be determined by measuring the optical reflection and absorption. In Fig. 4(c), we plot the dissipative (real) part of the optical conductivity of monolayer ScTe without and with the shear strain. One observes that without strain, the optical conductivity has a suppressed window from about 0.2 to 0.6 eV, as there is no available state for optical transition across the Fermi level in this frequency range. Around 1 eV, there appear two peaks at ~0.96 eV and ~1.13 eV, labeled as A and B in Fig. 4(c). By analyzing the band structure, we find that these two peaks correspond to locally parallel conduction and valence bands at the two locations marked in Fig. 4(d). They give an enhanced optical transition at these two particular frequencies. The applied shear strain produces noticeable changes in the optical conductivity. In Fig. 4(c), one can observe that the two peaks A and B are suppressed by the shear strain. In addition, due to the splitting of the nodal line by the strain, the intraband (Drude) contribution and the low-frequency interband contribution are enhanced. These contributions lead to the observed enhancement of the Drude peak. In other words, the splitting of the nodal line by strain can manifest as an enhancement of the Drude peak detected in optical absorption measurement.

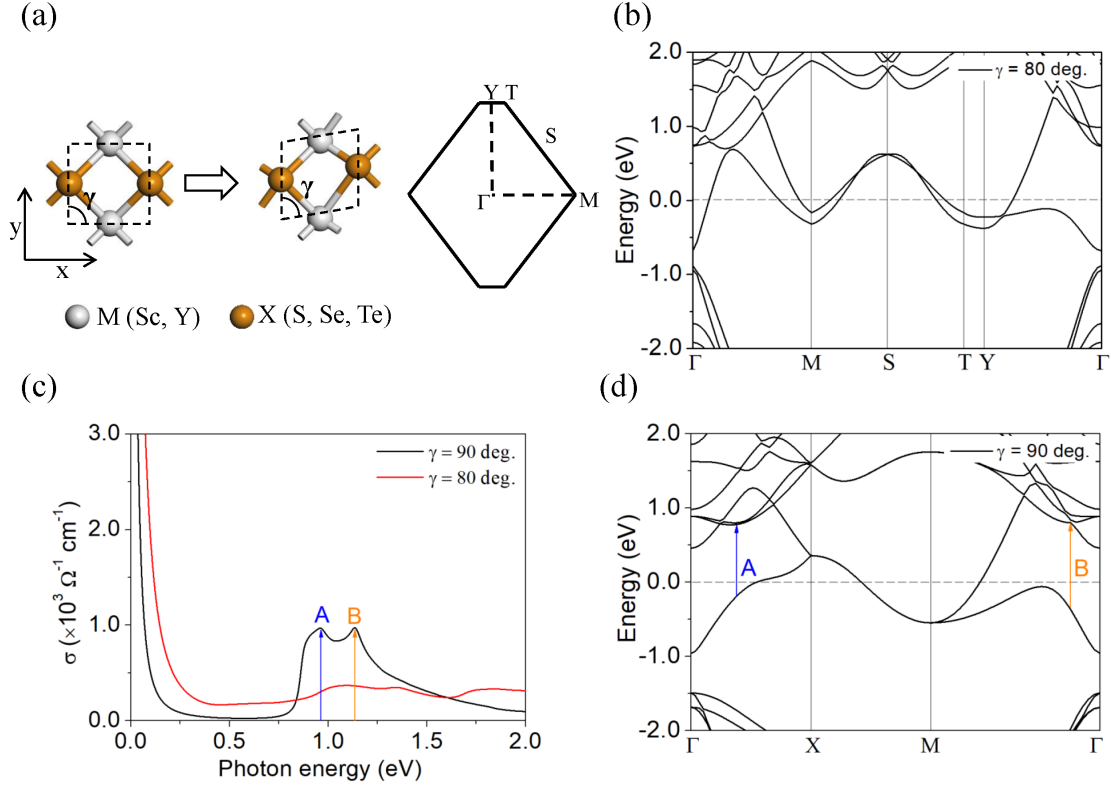

Fig. 4 (a) Left: schematic figure showing the shear strain considered here. Right: The BZ after applying the shear strain. (b) Calculated band structure of distorted (γ = 80 deg.) monolayer ScTe. (c) Optical conductivity of original (γ = 90 deg.) and distorted (γ = 80 deg.) monolayer ScTe. (d) Indicates the transitions that contribute to the peaks A and B in the optical conductivity for the undistorted monolayer ScTe.

## 4. Discussion and conclusion

In this work, we have predicted the stable 2D structures of *MX* (*M* = Sc, Y; *X* = S, Se, Te) family materials. To realize these materials, a possible approach is the bottom-up growth method, such as chemical vapor deposition or molecular beam epitaxy (MBE). For these approaches, one needs a suitable substrate for the material growth. Here, we screen several commonly used substrate materials with square lattice for the epitaxial growth of 2D films, including $SrTiO_3$(001) [57], $SiO_2$(001) [58], 3C-SiC [59], 6H-SiC(0001) [60] and $TiO_2$(001) [61], as listed in Table S1. From the comparison, we find that the lattice mismatches for $SrTiO_3$(001) and $TiO_2$(001) are minimal (typically < 6%), implying that the $SrTiO_3$(001) and $TiO_2$(001) could be suitable substrates for growing the *MX* monolayers in experiment.

To apply strains in experiment, one may choose a particular substrate for growth to achieve a small specific strain. For 2D materials, there are other well developed approaches. For example, one may transfer the fabricated sample to another substrate with trenches or holes, and use atomic force microscope tips to apply the strain [49]. Another approach is to transfer the sample to a flexible substrate, and apply strain via a beam bending apparatus [62].

In conclusion, based on the first-principles calculations, we discover a new family of 2D materials, the rare earth monochalcogenide *MX* (*M* = Sc, Y; *X* = S, Se, Te) monolayers. We demonstrate their stability and excellent flexibility with small Young's moduli and large Poisson ratios. These materials possess essential nonsymmorphic nodal lines on the boundary of the whole BZ around the Fermi level, which are protected by the combined operation of a screw rotation and the time-reversal symmetry in the absence of SOC. SOC opens a small gap on the nodal line, and the line evolves into stable 2D spin-orbit Dirac points. We demonstrate that the nodal line is robust against uniaxial and biaxial strains, but is lifted by the shear strain. We have also studied the optical conductivity of these materials, and show that splitting of the nodal line by strain can manifest in the optical absorption measurement. Our work offers a new platform for the study of 2D nonsymmorphic topological metal states. The excellent mechanical property and the interesting topological electronic property of these 2D materials may lead to potential applications in nanoscale devices.

**Acknowledgments**

This work was supported by the National Natural Science Foundation of China (No. 11602154 and 11604273), the Science and Technology Planning Project of Sichuan Province (NO. 2019YJ0157), and the Singapore Ministry of Education AcRF Tier 2 (MOE2019-T2-1-001). This work is also supported in part by the scholarship from the program of China Scholarships Council (No. 201906240171).